# Evidence for a Non-Expanding Universe: Surface Brightness Data From HUDF

Eric J. Lerner

*Lawrenceville Plasma Physics, Inc. Email:* `elerner@igc.org`

**Abstract.** Surface brightness data can distinguish between a Friedman-Robertson-Walker expanding universe and a non-expanding universe. For surface brightness measured in AB magnitudes per angular area, all FRW models, regardless of cosmological parameters, predict that surface brightness declines with redshift as $(z+1)^{-3}$, while any non-expanding model predicts that surface brightness is constant with distance and thus with z. High-z UV surface brightness data for galaxies from the Hubble Ultra Deep Field and low-z data from GALEX are used to test the predictions of these two models up to z=6. A preliminary analysis presented here of samples observed at the same at-galaxy wavelengths in the UV shows that surface brightness is constant, $\mu = kz^{0.026 \pm 0.15}$, consistent with the non-expanding model. This relationship holds if distance is linearly proportional to z at all redshifts, but seems insensitive to the particular choice of d-z relationship. Attempts to reconcile the data with FRW predictions by assuming that high-z galaxies have intrinsically higher surface brightness than low-z galaxies appear to face insurmountable problems. The intrinsic FUV surface brightness required by the FRW models for high-z galaxies exceeds the maximum FUV surface brightness of *any* low-z galaxy by as much as a factor of 40. Dust absorption appears to make such extremely high intrinsic FUV surface brightness physically impossible. If confirmed by further analysis, the impossibility of such high-$\mu$ galaxies would rule out all FRW expanding universe (big bang) models.

**Keywords:** surface brightness, cosmology, non-expanding universe, Tolman test.
**PACS:** 98.80.Es

## 1. SURFACE BRIGHTNESS TEST OF UNIVERSAL GEOMETRY

The Friedman-Robertson-Walker theory that is the basis for the conventional big bang model of the expanding universe makes the striking prediction that surface brightness of a given object decreases as $(z+1)^{-3}$, where z is redshift, if surface brightness is measured relative to AB magnitudes.[1] This factor is independent of all cosmological parameters. The relation is a consequence first of the time dilation factor of $(z+1)$ which reduces the number of photons received per unit time. In addition, there is an increase in angular size by a factor of $(z+1)$ due to the object being closer to the observer by this factor at the time the light was emitted. Thus the angular surface area is enlarged by $(z+1)^2$, leading to the overall $(z+1)^{-3}$ reduction in surface brightness. (The dependence of surface brightness on z is often expressed as $(z+1)^{-4}$, when surface brightness is measured as power per surface area, the additional factor coming from the loss of energy of each photon.)

In contrast, if the universe is not expanding, and the redshift-distance relationship is due to some other phenomena, surface brightness is independent of distance and thus of redshift. Apparent luminosity decreases as $d^{-2}$, where d is distance, and angular surface area decreases by the same factor, leaving surface brightness unchanged. Again this relation holds of all non-expanding models, regardless of the explanation of the Hubble relationship.

This test is closely related to the angular-size-redshift relationship, as in any non-expanding universe $\theta=k/d$, while in an FRW universe $\theta=k(z+1)/d$. However, the angular size test somewhat depends on the specific model or cosmological parameter set, which determines the relationship of d to the observable z.

The surface brightness test can be applied to galaxies, since, in the local universe, there is a relatively tight correlation of galaxy surface brightness, $\mu$, with galaxy luminosity, M. Therefore a comparison of the surface brightness of galaxies with the same luminosity at different redshifts can distinguish between the expanding and non-expanding models. The requirement that galaxies of the same luminosity be compared does introduce some dependence on specific models, as the determination of absolute luminosity depends on the relation of d and z, but as will be seen, model-independent conclusions can still be drawn.

Early studies of the $\theta$-z relationship using the linear extent of radio lobes in radio galaxies and quasars indicated a good agreement with the non-expanding model, assuming d=cz/H, as it is at low z.[2] But this was generally interpreted as evidence that these objects were physically smaller in the past, rather than as confirmation of a non-expanding universe. There was no evidence for this ad-hoc interpretation, however. An attempt was made to instead use compact radio sources as a standard ruler[3], and initially it was claimed that these objects followed the θ-z relationship predicted by FRW models. Later analysis [4] demonstrated that the large scatter in the size of the milliarcsecond sources and sample bias due to orientation made it almost impossible to draw conclusions from these objects.

Extensive optical and UV data on the size and surface brightness of high-z galaxies became available with the release of data from the Hubble Deep Field and more recently the Hubble Ultra Deep Field (HUDF). Preliminary analysis of this data indicated clearly that with ordinary galaxies, as with radio lobes, θ~1/z, consistent with the non-expanding model and in sharp contradiction to FRW predictions [5,6] Again these results were interpreted as size evolution of galaxies—as evidence that galaxies were smaller in the past, not as a test of the basic cosmological model.

Given the lack of data on high-z galaxies prior to 2003, use of the surface brightness test, also know as the Tolman test, was limited. Pahre, et al[7] in 1995 compared galaxy surface brightness from three cluster out to z=0.41 and Lubin and Sandage in 2001[8] used clusters up to z=0.9. In both cases, they concluded that surface brightness decreased with z but at a rate much less than that predicted than by FRW. Oddly, they concluded that this supported the FRW model, with evolution of surface brightness—brighter galaxies in the past—but excluded the non-expanding model with evolution—dimmer galaxies in the past. No justification was given for this argument. These studies were limited both by the modest redshifts involved and by the very limited samples, which were not necessarily unbiased, consisting only of galaxies in large clusters.

On the other hand, Andrews concludes that the surface brightness of brightest cluster galaxies is entirely compatible with the non-expanding universe hypothesis, although his data is also limited in redshift.[9] Also Jones and Disney concluded that the Hubble Deep Field data could not easily be explained by the conventional model[10]

The present study remedies the limitations of earlier studies. It is the first to use the high-z data from HUDF, which extends up to z=6, combined with very extensive low-z data from GALEX, with redshifts provided by the Sloan Digital Sky Survey (SDSS). By using high-z data, we can much more decisively distinguish between the FRW and non-expanding predictions, since at z=6, the surface brightness predictions of the two models differ by a factor of $7^3$ or 343. In addition, as will be shown, the redshifts are sufficiently great that the evolutionary modification of the FRW predictions can be tested in a definitive manner.

## 2. TEST OF EUCLIDEAN NON-EXPANDING UNIVERSE

### 2.1 Data Description

For this study, the high-z data, derived from the HUDF images, consists of a number of data sets. The highest-z set consists of i-drop galaxies, which are galaxies selected by photometric color to have z~5.5-6.5. These have non-detection in the 606 nm v band and i'-z'>1.4, where i' is AB apparent magnitude in the 775 nm band and z' is the same in the 850 nm band. This data[11] consists just of 850 nm half-light radii and 850 nm AB magnitudes for each of 38 galaxies. Similarly, the v-drop galaxies, selected in a similar manner, have average redshifts of 4.9 and the same data for each of 156 galaxies. The third set was created by collating together a set of 1727 galaxies observed by NICMOS, which had photometric redshifts, with 7016 galaxies in the UDF-z catalog, which have 850 nm AB magnitudes and half-light radii, producing a set of 523 galaxies in both data sets[12,13]. This third NICMOS-HUDF data set also contains other data on each galaxy, including ellipticity, and size and magnitude in other bands. This third data set was used for galaxies having photometric redshift from 2.5 to 4. For 2.5<z<6, the 850 nm bands correspond to at-galaxy wavelengths of 242-121 nm, in other words from near to far UV. (The term "at-galaxy" is used, rather than "rest frame" since it applies to both expanding and non-expanding models.)

The low-z comparison data set consists of galaxies that are both in the GALEX data sets and have good spectroscopic redshifts from SDSS. Two GALEX data sets are used—the All Sky Imaging Survey (AIS) and the Medium Imaging Survey (MIS), which has longer exposure times. The AIS+SDSS set consists of 28,394 galaxies while the MIS+SDSS sample consists of 10,150 galaxies. These data sets contain AB magnitudes and half light radii at FUV (155 nm) and NUV (230 nm) bands, redshifts, as well as data for the SDSS optical bands. All the data will be made available online.

## 2.2 Model to be Tested and Sample Selection

The first surface brightness test is of the non-expanding universe hypothesis. Although all non-expanding models predict a constant surface brightness, the dependence of surface brightness, $\mu$, on galaxy luminosity, M, necessitates selecting a particular relationship between z and d. Figure 1 shows the $\mu$–M relationship for low-z galaxies observed in the FUV band. For clarity, M is defined as m-5 Log z, where m is the measured AB magnitude. These values can be converted into conventional absolute magnitudes $M_{abs}$= M-43.0. For comparison, in these units, the luminosity of the Milky Way is 23.3. Surface brightness is in AB magnitudes per $arcsec^2$, and again the Milky Way's surface brightness is about 24 mag/ $arcsec^2$. As mentioned in section 1, given the $\mu$–M relationship, galaxies of the same M must be compared for surface brightness. Since determining M requires knowing d, which cannot be directly measured at high z, a specific relationship of z with d must be assumed.

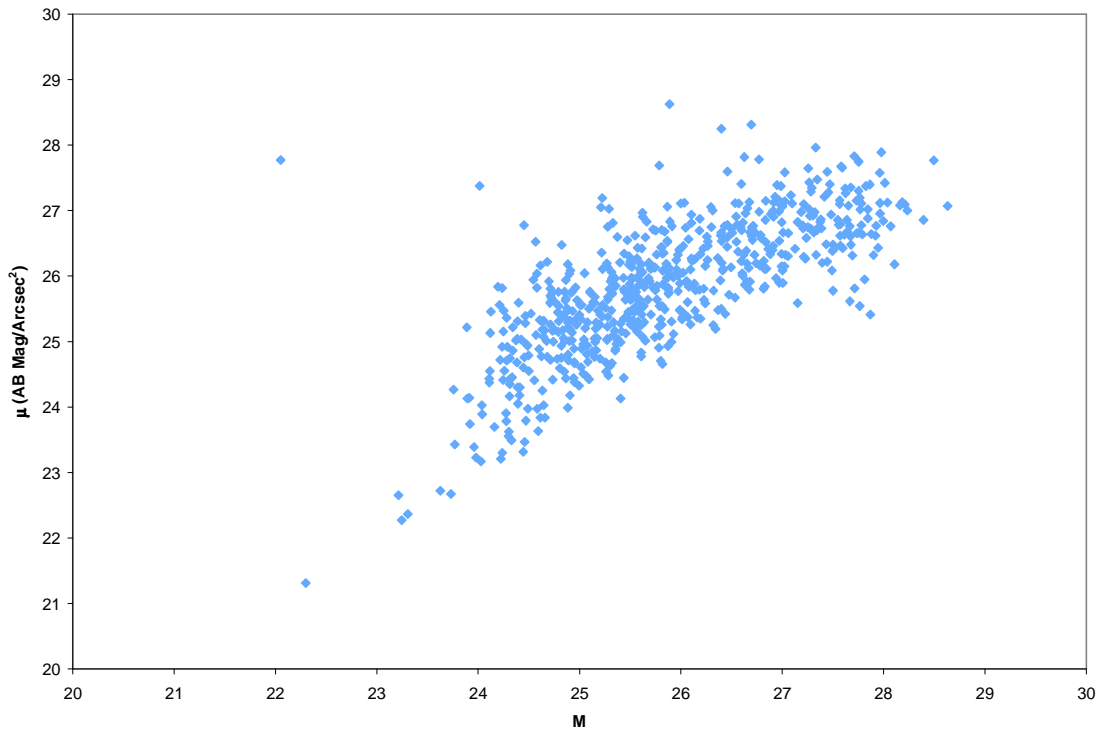

**FIGURE 1.** Surface brightness $\mu$ vs galaxy absolute luminosity M=m-5Log z, where m is FUV apparent magnitude in AB magnitudes for GALEX MIS galaxies with SDSS redshift for the range 0.135<z<0.16.

Here it is assumed that d=cz/H for all z, as it is observed to be at low z, where d can be independently measured. This assumption is chosen because it requires no specific hypothesis of the mechanism that generates the redshift, but merely extrapolates the known behavior of the redshift-distance relation that has been observed. The effect of alternative d-z relationships will be briefly examined in section 4. In addition, we assume no evolution in the $\mu$–M relationship.

In selecting specific samples of high-z and low-z galaxies to compare, several considerations are important, in addition to selecting for similar ranges of M. First, $\mu$ also varies with wavelength, $\lambda$, so the at-galaxy $\lambda$ of the high and low-z samples must be matched a closely as possible. Second, the sample's low-size cutoffs, the smallest galaxy that can be observed, must be matched. The HUDF sample has a 0.03" pixel size, but in practice the minimum observable galaxy radius is 0.06". GALEX has a 1.5" pixel size and due to different image processing, the smallest galaxies in the AIS sample have radii of 1.5" and those in MIS 1.8".To test the non-expanding hypothesis, we must compare samples that have the same cutoff in *physical* size, so we need samples whose cutoff have the same θd or θz. Note that sample selection depends here on the z-d relationship assumed. Third, we require that both high and low–z samples have adequate size for accurate statistics.

Finally, we have to avoid selection effects that imposes a large-size cutoff. The selection criteria and methods of observation of SDSS eliminate some of the galaxies with largest angular dimensions, a selection that varies with apparent magnitude. This effect, which can artificially truncate the low-z surface brightness distribution, is less significant with the deeper-exposure MIS sample than with AIS. As will be shown, with proper selection of samples, this effect can be avoided.

The first high-z data set is the i-drop set centered at z=6. The 850 nm z-band corresponds to an at-galaxy λ of 121 nm. As can be seen from figure 2, the i-drop galaxies are concentrated into a narrow range of M, so the range from 23-24 is selected, which gives N=25 galaxies. The low-size cutoff is θz=0.36, which is a physical radius of 7 kpc for H=75 km/sec/Mpc with d=cz/H. At low-z, the FUV 155 nm band corresponds to an at-galaxy λ of 121 nm for z=.281. At this z, there are too few galaxies in the MIS sample but the AIS sample meets all our criteria. To match the proportional range in z of the high-z sample of about 5.5-6.5, we select a low-z range of .263-0.3. Within this range the low-z sample contains 47 galaxies, an adequate number. The low-size cutoff is 8.2 kpc. The large-size cutoff is not a problem, since this sample contains no galaxies larger than 4.5", yet within the sample range of m, galaxies with θ as large as 10" are contained in the overall AIS-SDSS sample. To match the samples as closely as possible, we also eliminate the brightest galaxy in the low-z sample, leaving 46, so that average M in the low-z sample is 23.50 and in the high-z sample are 23.49.

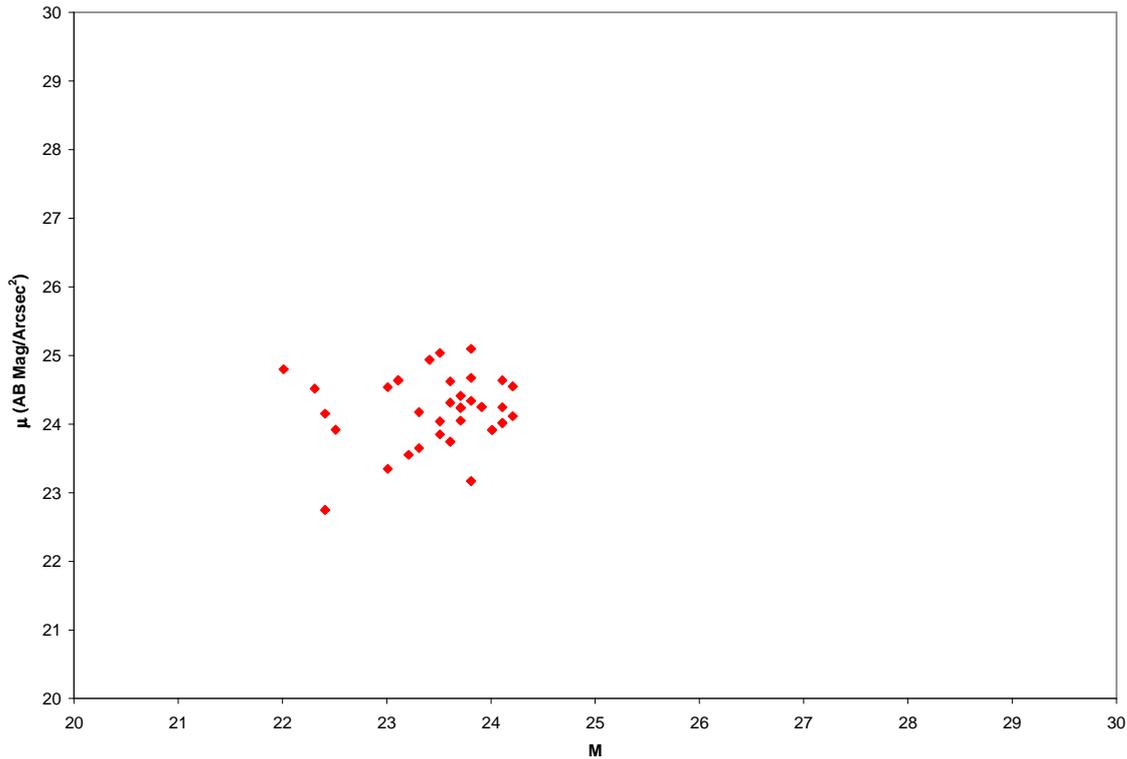

**FIGURE 2.** Surface brightness μ vs galaxy absolute luminosity M=m-5Log z, where m is 850 nm (z-band) apparent magnitude in AB magnitudes for HUDF i-drop galaxies with photometric z~6.

The second high-z data set is the v-drop set centered at z=4.9. The 850 nm z-band corresponds to an at-galaxy λ of 144 nm. This data set is shown in Figure 3. On the one hand, it's desirable to choose galaxies that are close in M to that of the i-drop sample, so that similar galaxies can be compared at different z. On the other hand, it is important to maximize the number of galaxies in the sample. So here the range of M from 23.5-24.5 is selected, giving N=45 galaxies. The low-size cutoff is θz=0.29 or 5.7 kpc. At low-z, the FUV 155 nm band corresponds to an at-galaxy λ of 144 nm for z=.076. However at this z there are too few galaxies in either AIS or MIS sample and the low-size cutoff at 2.7 kpc is too small. To attain adequate sample size, we select the redshift range 0.135<z<0.16 in the MIS sample, which has a low-size cutoff of 5.1 kpc. This sample has an at-galaxy λ centered at 135 nm. As can be seen from Fig. 1, the 23.5-24.5 range is necessitated to get adequate numbers in the low-z sample. There are 69 galaxies in this range, of which none are smaller than the high-z cutoff of 5.7 kpc. The large-size cutoff is not a problem, since this sample contains no galaxies larger than 17", yet within the sample range of m, galaxies with θ as large as 24" are contained in the overall AIS-SDSS sample. To match the samples as closely as possible, we also eliminate the 12 dimmest galaxies in the low-z sample, leaving a total of 57, so that average M in both the low-z sample and in the high-z sample is 24.19.

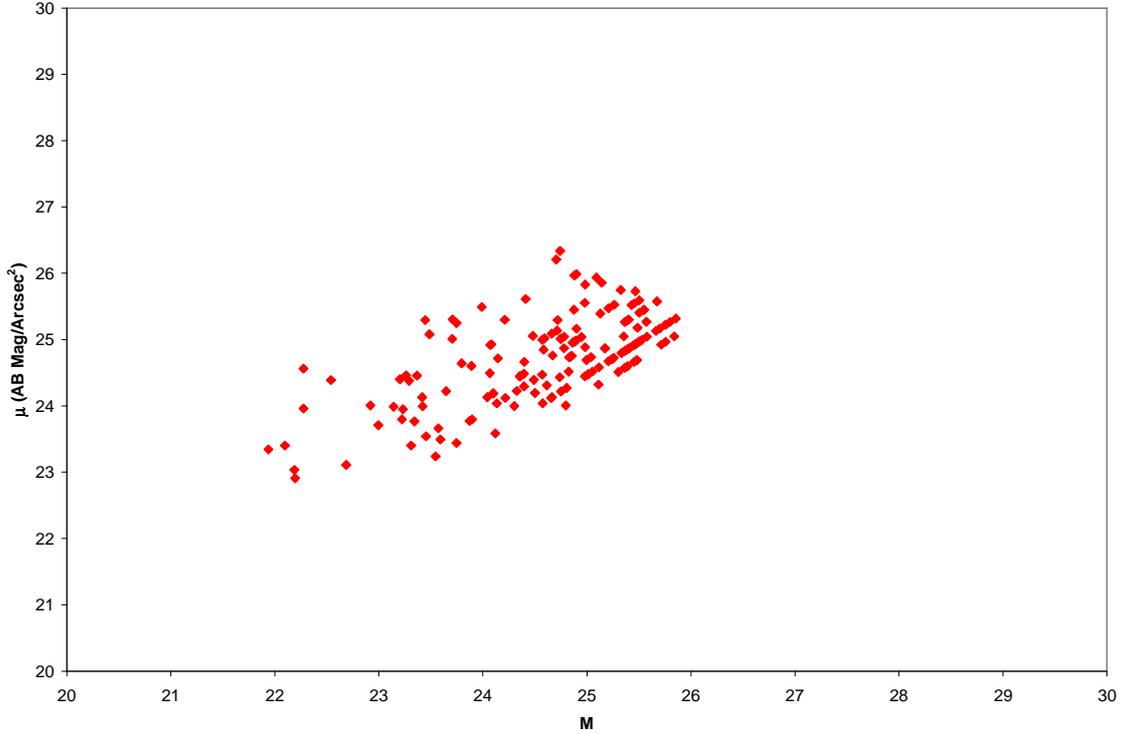

**FIGURE 3.** Surface brightness μ vs galaxy absolute luminosity M=m-5Log z, where m is 850 nm (z-band) apparent magnitude in AB magnitudes for HUDF v-drop galaxies with photometric z~4.9.

The last high-z data set consists of those galaxies with z=2.5-4.0 in the NICMOS-HUDF data. This range was chosen to lie below the range of the v-drops but to still have sufficiently high z so that the 850 nm band could is compared with GALEX data. The 850 nm band corresponds to an at-galaxy λ of 170-242 nm with an average of 200 nm. The data set is shown in Figure 4. Again the range of M from 23.5-24.5 is selected, giving N=32. The low-size cutoff averages θz=0.20 or 3.8 kpc. At low-z, the NUV 230 nm band corresponds to an at-galaxy λ of 200 nm for z=0.15. We get adequate sample size for z=0.11-0.13 with a low-size cutoff of 4.2 kpc and λ of 205 nm. The large-size cutoff is not a problem, since this sample contains no galaxies larger than 13", yet within the sample range of m, galaxies with θ as large as 20" are contained in the overall AIS-SDSS sample. For M=23.5-24.5, N=96, and again the 20 dimmest galaxies are eliminated to create a final sample of 76 galaxies with average M of 24.06, identical with the high-z sample.

It should be noted that the limiting surface brightness in the HUDF samples is 26.5 mag/arcsec$^2$, which is considerably dimmer than the dimmest galaxies in the selected samples, so does not bias the sample. That is, if there were galaxies in this luminosity range with lower surface brightness, they would have been observed.

## 2.3 Data Analysis and Test of Model

The comparison of the high-z and low-z pairs of samples is shown in Table 1 and Figure 5. Note that in Table 1 the surface brightness is in mag/arsec$^2$, while Figure 5 presents the logarithm of the difference in surface brightness from the low-z samples. In this table, the subscripts H and L refer to the high-z and low-z samples respectively and $\Delta \mu = <\mu>_H - <\mu>_L$. If we require that a linear fit to the three high-z points in Figure 5 also pass through the low-z point (by definition) we obtain the following best fit

$$\text{Log } \mu = (0.024 \pm 0.15) \log z + 0.022 \qquad (1)$$

This is an excellent confirmation of the non-expanding prediction that μ is independent of z. The surface brightness data is thus entirely consistent with the hypothesis that the universe is not expanding and that d=cz/H at all z.

If we look at the individual comparisons in Table 1 we see that in all three cases the differences in μ are less than 1 magnitude, but in two of the three cases they are statistically significant. Such shifts could either be due to evolutionary effects—changes due to the epoch of creation of the galaxies, or to environmental effects—changes due to the density or number of clusters in the relatively small volumes probed at different z. While these two hypotheses cannot be distinguished in the data, it is important to see if the trend toward increasing μ at lower z continues or is merely a fluctuation based on the specific volumes sampled.

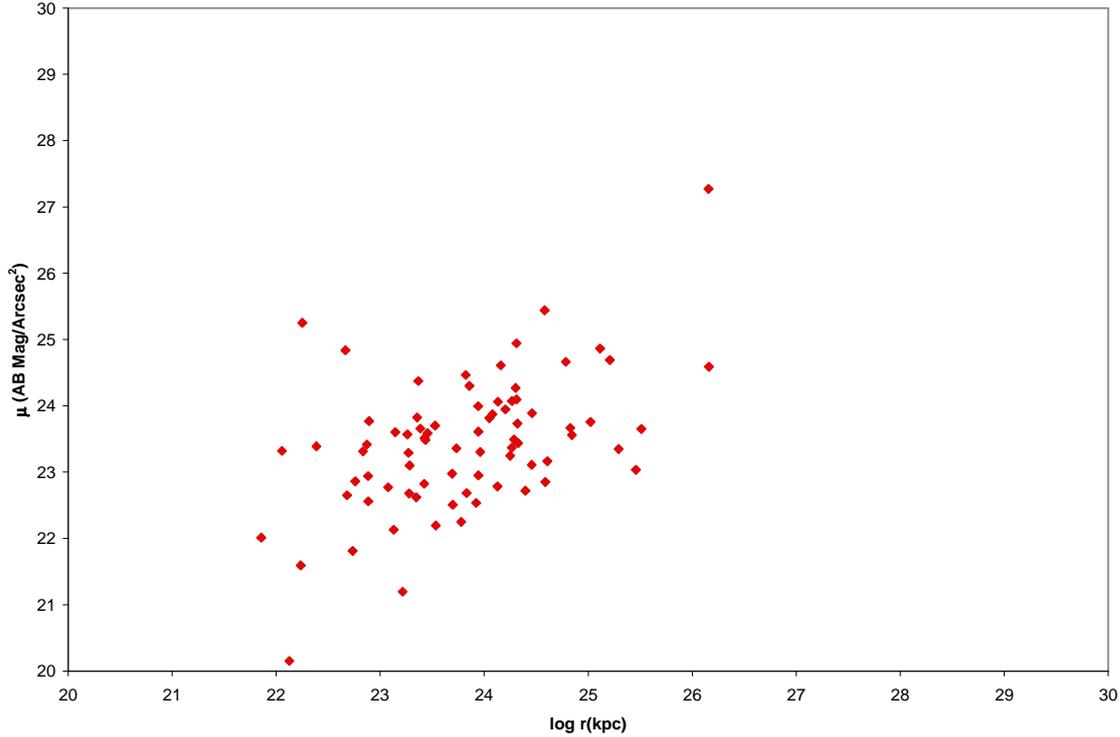

**FIGURE 4.** Surface brightness μ vs galaxy absolute luminosity M=m-5Log z, where m is 850 nm (z-band) apparent magnitude in AB magnitudes for HUDF galaxies with photometric <2.5z<4.

**TABLE 1** Non-Expanding Universe Comparison

| Pair number | $N_H$ | $<z>_H$ | $<\mu>_H$ | $\sigma_H$ | $N_L$ | $<z>_L$ | $<\mu>_L$ | $\sigma_L$ | $\Delta \mu$ | $\Delta \mu /\sigma_H$ |
|---|---|---|---|---|---|---|---|---|---|---|
| Pair 1 | 25 | 6.0 | 24.22 | 0.50 | 46 | 0.281 | 23.98 | 0.63 | 0.24 | 2.4 |
| Pair 2 | 45 | 4.9 | 24.45 | 0.54 | 57 | 0.148 | 24.43 | 0.88 | 0.02 | 0.25 |
| Pair 3 | 32 | 3.2 | 23.53 | 0.70 | 76 | 0.12 | 24.34 | 0.75 | -0.81 | -6.5 |

To distinguish these two possibilities, it is important to probe surface brightness at intermediate z. While this is a major goal for future work, one example near z=1 is included here. The data available at present does not allow exactly the same analysis as at higher z, but a similar analysis is possible. To compare NUV at low redshift with z~1 data, we need surface brightness data from the 435 nm ACS band. But half-light radii at this band are not available in this data base. Instead, a different measure of galaxy radius, full-width-half-maximum, is available at both 435 nm and 850 nm. We can therefore determine the ratio of surface brightness at 435 nm to that at 850 nm in a consistent manner. Since the half-light radius data is available at 850 nm, we can also determine the ratio of surface brightness for a given galaxy measured by FWHM and by half light radius at 850 nm. If we assume that FWHM/HLR is the same at the two wavelengths, we can then determine the 435 surface brightness as measured by HLR.

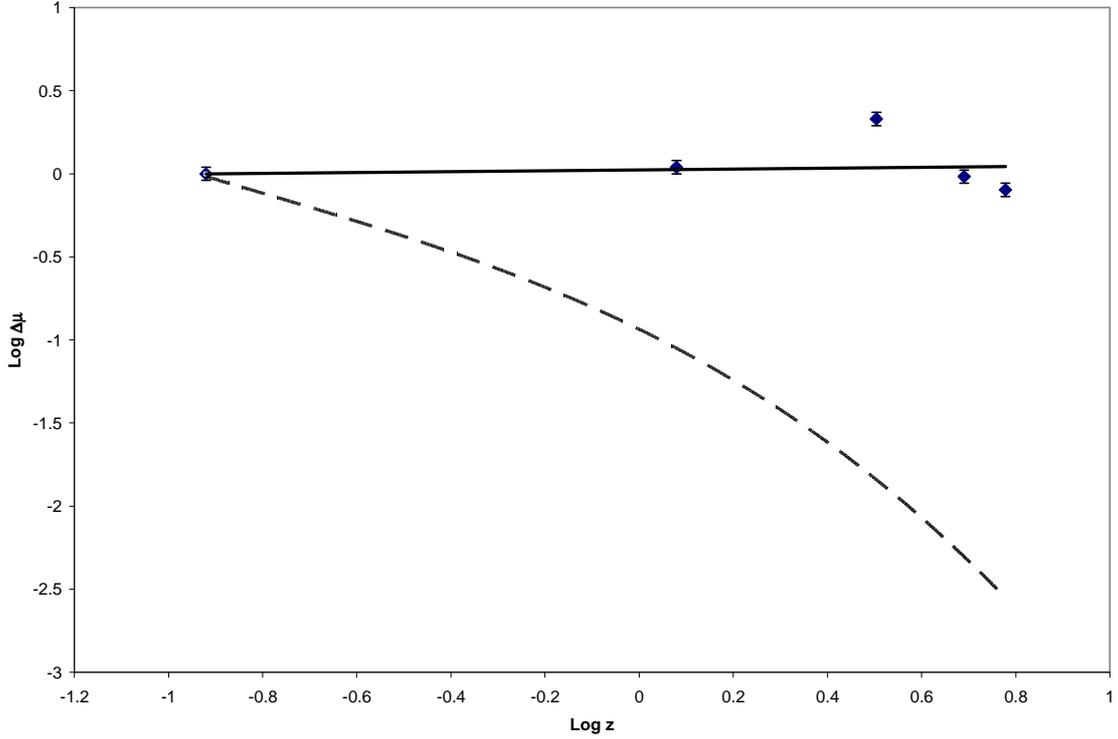

**FIGURE 5.** The logarithm of the difference in surface brightness between high-z and low-z galaxy samples $\Delta\mu$ plotted against logarithm z. $\Delta\mu$ is defined so that positive values correspond to the high-z samples having higher surface brightness. The three data points at highest z correspond to the values from Table 1, while the fourth point is obtained slightly differently (see text). The fifth point at low z is set at 0 by definition .The solid line is a linear fit to the data, constrained to pass through the low-z data point. It is indistinguishable from the prediction of the non-expanding universe hypothesis, which is the x-axis. The dashed line shows the prediction of the expanding-universe FRW model, without assuming evolution. The line curves because it is plotted against log z not log (z+1).

$$\mu_{HLR,435} = \mu_{FWHM,435} (\mu_{HLR,850} / \mu_{FWHM,850}) \qquad (2)$$

To compare 435 nm data with the NUV sample, we select an intermidiate z sample z=0.9-1.3 with the at- galaxy $\lambda$ of 205 nm and a low-size cutoff of 1.3 kpc. For comparison we select low-z sample with z=0.07-0.09 with $\lambda$ of 213 nm and a low-size cutoff of 2.8 kpc. To obtain sufficient galaxies in the high-z sample we shift to a lower-luminosity range of M=24.5-25.5, yielding 15 galaxies, of which one is eliminated for having a radius less than the low-z cutoff of 2.8 kpc. There is a large comparison sample of 280 galaxies at low z, of which the dimmest 20 are eliminated to produce identical average M of 25.03 in both comparison samples. The average $\mu$ for these two comparison samples are very close: 24.27 mag/arcsec$^2$ for the low-z sample and 24.17 mag/arcsec$^2$ for the high-z sample. (This comparison is included as a point in Figure 5.) With this point included, the linear fit to the data is:

$$\text{Log } \mu = (0.026 \pm 0.15) \log z + 0.024 \qquad (3)$$

Thus there is no real trend and surface brightness is indeed independent of z over the entire range of z observed. Again the data are entirely consistent with a non-expanding universe.

This conclusion contradicts the earlier work of Lubin and Sandage. However, there are many limitations of that earlier work as compared with this one. Lubin and Sandage did not compare galaxies observed at the same at-galaxy wavelengths at low and high redshift as this study does. Instead they used a very involved evolutionary k-correction scheme, with many adjustable assumptions and parameters, to "correct" observed high-z surface brightness. This process introduced unknown and possibly large errors into the calculation, In addition the samples were very small and the range of redshifts, up to 0.9, much smaller than that of the present study (up to 6). Finally, they assumed a different relationship of z and d (d=c ln (1+z)/H) to test the non-expanding hypothesis, although, as we shall see, this probably does not have a major effect.

## 3. TEST OF FRW EXPANDING UNIVERSE

To test the FRW expanding universe hypothesis, somewhat different comparison samples are needed. This hypothesis assumes that angular radii for a given galaxy are increased at high-z, so the assumed physical size of the high-z galaxies is much smaller for an observed angular size. The exact factor depends on the assumed distance, which in turns depends on the exact model, but for the "consensus model" radii are reduced by a factor of 20 at z=6 relative to the non-expanding model and by a factor of 11 at z=3.2. This means that the comparison samples at low z must have a much lower low-size cutoff and therefore must be much closer. For example, since the low-size cutoff remains roughly constant from z=2.5 to z=6 at around 0.35kpc, the comparison sample has to have 0.01<z<0.02.

For the z=6 and z=4.9 samples there are no suitable comparisons, as the range of M in the low–z samples barely overlap with that of the high-z sample. The high-z samples do not contain enough low-luminosity galaxies for comparison. However, a comparison can be made of the low-z sample with the 2.5<z<4 sample.

The 0.01<z<0.02 sample in the NUV has an at-galaxy $\lambda$ of 227 nm as compared with the 200 nm average for the high-z sample. Both samples have the same low-size cutoff of 0.35 kpc (assuming FRW geometry). While there is still no real overlap in the absolute magnitudes, both samples have sufficient spread in M to derive a $\mu$-M correlation. For the low z sample, the linear best fit is $\mu$ =0.597M+7.89 while for the high-z sample it is $\mu$ =0.551M+10.31. If we compare the two correlations at M=25 we find that, unsurprisingly, there is now a considerable difference in surface brightness—average surface brightness is 22.81 mag/arcsec$^2$ at low z and 24.09 mag/arcsec$^2$ at high z, a difference of 1.29 mag/arcsec$^2$. (This does not contradict the non-expanding results, since with the non-expanding assumption this low-z sample includes much smaller galaxies than could be observed at high-z, so is not a good comparison.) Since the slopes are nearly the same, comparison at other m are very similar. For example at M=24 dm= 1.33 mag/arcsec$^2$ .A correction of 0.08 can be applied to the high-z $\mu$ to compensate for the shift in wavelength, but this is not very significant.

Ignoring this correction, we can compare the change in surface brightness with that predicted by FRW. The predicted change in $\mu$ is just 7.5 log (1+z) = 4.67 mag/arcsec$^2$ for z=3.2. The actual observed change (using FRW assumptions) is 1.29 mag/arcsec$^2$, a clear conflict with prediction. Put another way, the observed dependency of surface brightness on redshift, assuming FRW, is $(z+1)^{0.83}$, not $(z+1)^3$.

### 3.1 Test of the Evolution Hypothesis

As mentioned above, many authors, including Lubin and Sandage, have noted the difference between surface brightness observations and the clear predictions of the FRW model. They have consistently explained this discrepancy as a consequence of galaxy evolution. That is, they assume that galaxies of a given luminosity were smaller and thus had brighter $\mu$ in the past than do current galaxies. Equivalent, galaxies of a given size had much higher luminosity in the past then at present. One could ask why galaxies at high-z are so much brighter, when young galaxies observed at low-z are not. However, at most wavelengths, such as in the optical, there is not a clear way to test the evolution hypothesis.

Fortunately, the situation is different at UV wavelengths. UV radiation is generated by hot, massive short-lived stars and is strongly absorbed by dust. The dust in turn is created by the supernovae generated as the massive stars reach the end of their short lifetimes. The combination of these effects means that there is an upper limit on the UV surface brightness—as the surface density of hot bright stars and thus supernovae increases, eventually so much dust is produced as to absorb all the UV except that from a thin surface layer. Further increase in surface density of hot bright stars beyond this point just produces more dust, and a thinner surface layer, not an increase in UV surface brightness.

To simplify the situation somewhat, assume that a population of N hot bright stars produces E joules of FUV radiation and D kg of dust during their lifetimes of $\tau$ seconds. The dust has an absorption cross section of C m$^2$/kg and the density of the stars is n/kpc$^3$. The absorption length for FUV radiation is then

$$L = 9 \times 10^{38} \, N/DCn \text{ kpc} \quad (4)$$

If the galaxy thickness T< L in the direction of the line of sight, the FUV surface brightness in watts/ m$^2$ is just

$$\mu' = 1.1 \times 10^{-39} \, TEn/N\tau \quad (5)$$

But if T>L, in other words absorption is substantial, then little FUV is emitted from a greater thickness than L so, approximately,

$$\mu' = 1.1 \times 10^{-39} \, LEn/N\tau = E/DC\tau \quad (6)$$

In other words, the maximum FUV surface brightness depends only on the ratio of FUV power to dust production for the stellar population, not on the density of stars in the galaxy. A more sophisticated model would include the influence of galactic winds in clearing away the dust, but the general conclusion stands.

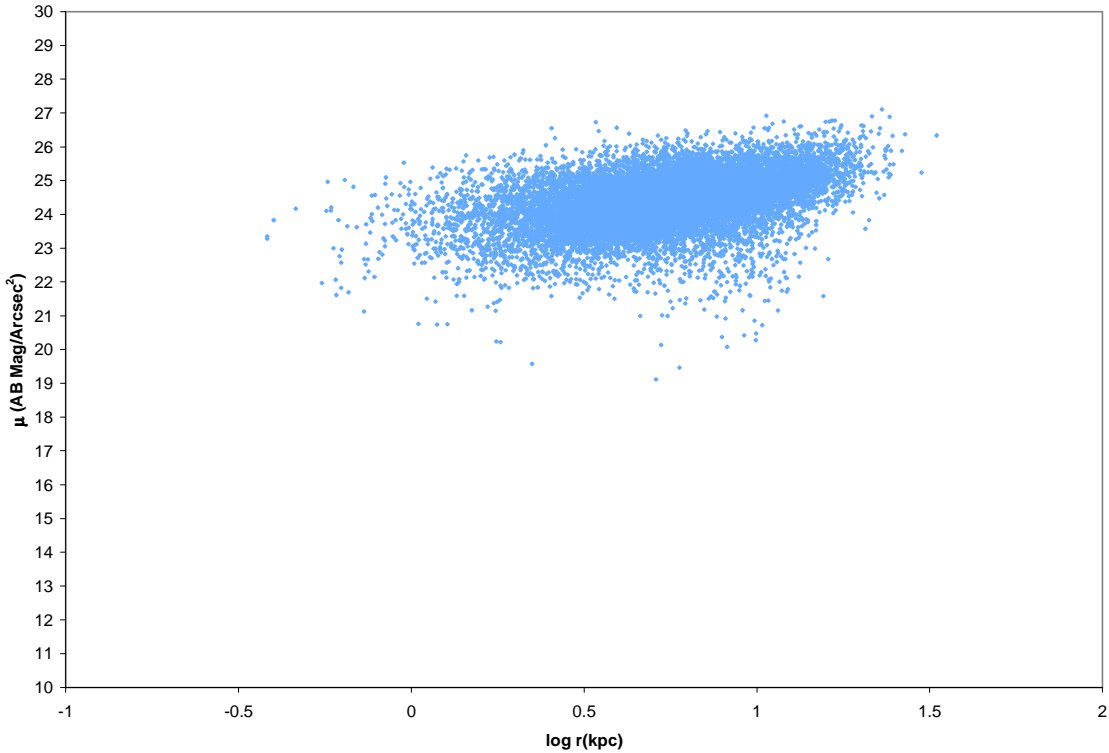

**FIGURE 6.** Observed surface brightness against galaxy radius for GALEX-SDSS galaxies with 0.01<z<0.25. Radii are calculated based on d=cz/H and no expansion, but due to the low z, the results are very similar if FRW geometry is assumed.

We can now use this maximum UV surface brightness to test the evolutionary hypothesis. The GALEX-SDSS data set can be used to determine the maximum FUV and NUV surface brightness of galaxies. This in turn can be compared with the intrinsic surface brightness that must be hypothesized for high-z galaxies in the FRW model. This is the observed surface brightness multiplied by $(z+1)^3$. The test is to see if the hypothetical surface brightnesses exceed the maximum surface brightness observed at low-z and are thus physically implausible.

Figure 6 shows FUV surface brightness vs. galaxy radius in kpc for the redshift range 0.01-0.25 including 13,858 galaxies. The highest surface brightness is 19.1 mag/arcsec$^2$ and only 18 galaxies have µ brighter than 21 mag/arcsec$^2$. So we can tentatively take 19.1 mag/arcsec$^2$ as the maximum possible FUV surface brightness.

The SDSS survey has a surface brightness upper limit in the r-band of 17.7 mag/arcsec$^2$, which could affect the observation of the UV distribution. However, examining GALEX subsamples that do not have SDSS matches, we find the same surface brightness upper limit. GALEX's own instrumental upper limit is far higher, with linear response to about 14.5 mag/arcsec$^2$ and a cutoff at around 11 mag/arcsec$^2$.

If we look at the galaxies that are known to have the maximum rate of star formation and the maximum overall surface brightness, the ultra-luminous infrared galaxies (ULIRGs), we can see the limiting effect of absorption clearly. The sample of ultra-luminous galaxies observed in the UV by Goldader et al [14] provides IR and FUV surface brightness at various apertures for a set of nine galaxies. Figure 7 shows FUV µ vs IR µ for both maximum and minimum apertures for each galaxy. Over a range of 1,000 in IR µ there is no trend in FUV µ. All but one of this much smaller sample of very luminous galaxies has FUV µ dimmer than 21 mag/arcsec$^2$ and the brightest one has a µ of 19.36 mag/arcsec$^2$. So 19.1 mag/arcsec$^2$ can be confidently used as a maximum FUV surface brightness.

To compare these limits with the evolutionary hypothesis, the brightest, highest-z galaxies are selected. This is the sample of the NICOMS-HUDF data set with z=3.5-6. This sample was not chosen for the earlier surface brightness comparison because it was too small, but is suitable for this comparison. Figure 8 shows the hypothetical at-galaxy surface brightness of this sample, derived by multiplying the observed surface brightness by $(z+1)^3$ or µ''= µ-7.5 log (z+1). The low-z data is included for comparison. As is evident the hypothetical galaxies lie almost entirely outside the envelop of the real low-z galaxies. All but one of the 29 hypothetical galaxies is brighter than 19.1 mag/arcsec$^2$. Eighteen of the galaxies exceed the low-z maximum µ of 19.1 mag/arcsec$^2$ by more than a factor of 2. Four of the galaxies exceed the maximum by a factor of 10. The brightest hypothetical galaxy exceeds the maximum low-z µ by a factor of 42.

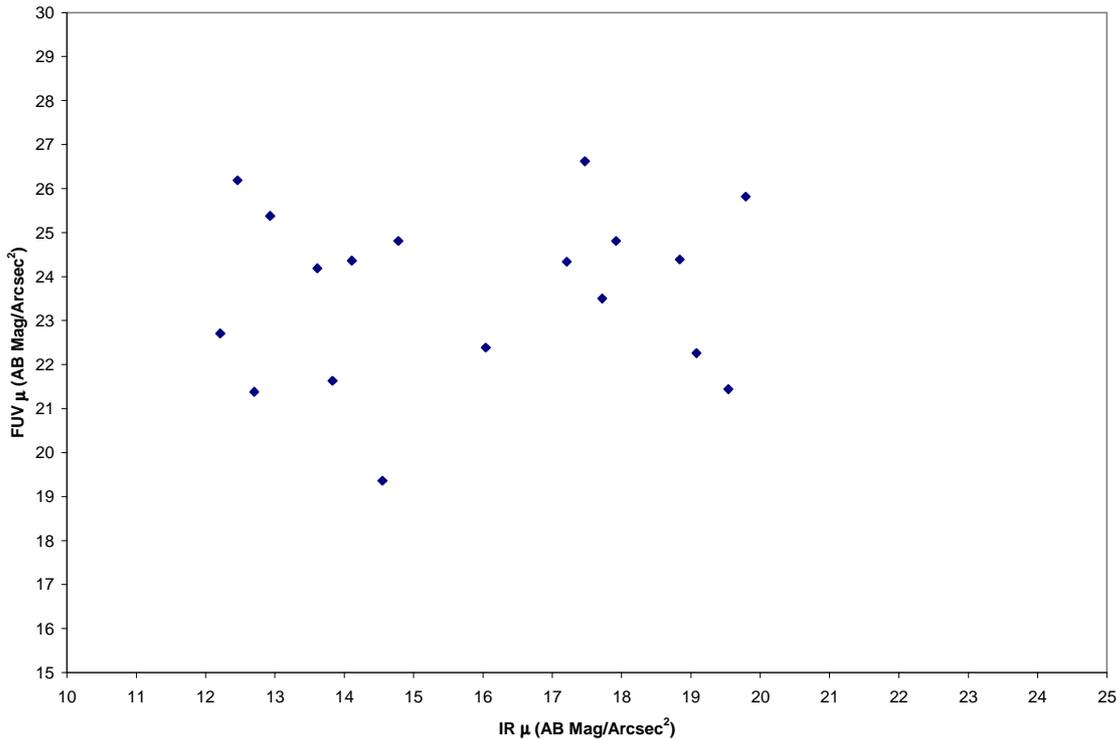

**FIGURE 7.** FUV vs IR surface brightness for nine ULIRG galaxies.

It is exceedingly difficult to see how these hypothetical galaxies could actually exist. We can compare them with ULIRGs that have similar total surface brightness since the hypothetical galaxies must have at least this much total output, even if all the energy is in the FUV and none is absorbed by dust. Figure 9 shows the ratio of IR to FUV emission in magnitudes vs. IR surface brightness. There is an almost linear dependence of absorption on IR surface brightness, as is expected from eq. (4). To reduce absorption to less than 0.5 magnitudes, as would be needed to achieve the FUV surface brightnesses of the hypothetical galaxies, the ratio of dust production to UV power would have to be decreased by at least a factor of ten.

Observations of high-z galaxies at other wavelengths indicate that there is very little dust absorption, in strong contrast to the large absorption of the ULIRGs[15]. This would of course be easy to explain if the high-z galaxies actually had their observed μ, as they would in a non-expanding universe. They would then all have μ>21 mag/arcsec$^2$ and would be expected to have small absorption.

It can be argued that these observed galaxies are merely small "hot spots" within much dimmer and larger galaxies. However, this does not at all solve the problem. The ULIRG sample includes narrow-aperture measurements that focus on the brightest regions of the galaxies, and the GALEX survey does not distinguish between complete galaxies and star-forming clusters within galaxies, yet in neither of these samples is the maximum source brightness of 19.1 mag/arcsec$^2$ exceeded for any region even as small as 30 pc in radius.

How could the hypothetical galaxies possibly produce ten times less dust per unit FUV power than any contemporary galaxy? One possibility is that the high-z galaxies contain a population of very large stars, more than one hundred times as massive as the Sun, with much shorter τ, leading to higher power for the same amount of dust. However, such very hot stars will produce substantially more radiation at FUV than at NUV at-galaxy wavelengths. Since we have 1600 nm as well 850 nm data, we can test this hypothesis. We find that of the six hypothetical galaxies with the brightest μ", only one is substantially (more than 0.2 mag) brighter at 850 nm than 1600 nm. So for the other five galaxies, the idea of ultra-massive, ultra-bright stars is not a likely solution. It should also be noted that in the low-z sample, there are many galaxies brighter in FUV than NUV, yet none of them have μ<20 mag/arcsec$^2$, so even a population of very massive hot stars does not lead to excessively high UV surface brightness.

Nor are energetic galactic winds that clear out the dust far faster than for low-z galaxies a very plausible explanation either. Such winds would have to have at least ten times higher velocity and thus 100 times more power than low-z winds per unit mass. It seems entirely implausible that the efficiency of galactic wind production should be at least 100 times greater for high-z galaxies, which after all contain stars very similar to, if not identical to, those in low-z galaxies. While a more systematic examination of possible explanations is required, the tentative conclusion in unavoidable that at least the most extreme of these hypothetical galaxies cannot really exist.

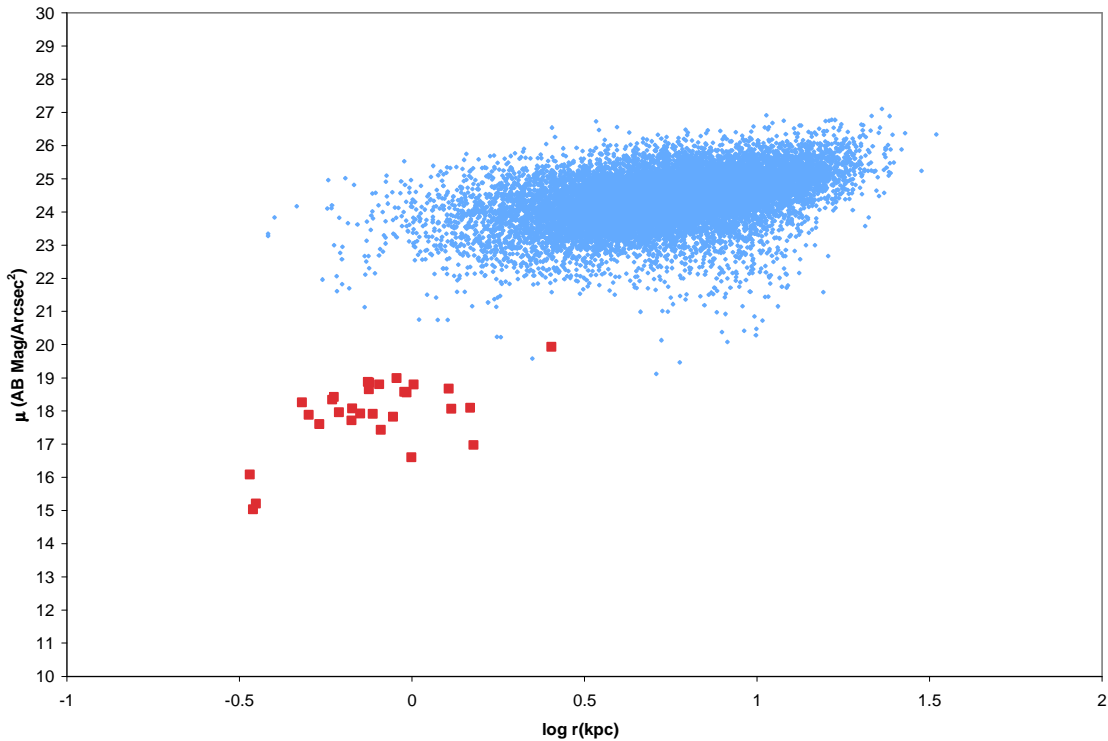

**FIGURE 8.** As in Figure 6, but with the addition of hypothetical high-z galaxies (large squares), with surface brightness and radii as required by FRW expanding universe model, based on observations of galaxies with photometric redshift 4<z<6. Note that the hypothetical galaxies lie almost wholly outside the envelope of the real low-z galaxies (small dots).

But if *any* of these galaxies are excluded as impossible, this means that the $(z+1)^3$ surface brightness scaling cannot be valid. This in turn means that the FRW expanding universe model, in all its varieties, is invalidated.

By contrast, if this same sample is analyzed in the non-expanding model context, all the galaxies have $\mu > \mu_{max}$ and none fall outside the range of parameters defined by the low-z sample.

Of course, another drawback of the evolutionary approach is that the excellent fit of the data to the non-expanding model then can only be interpreted as a coincidence.

## 4. DISCUSSION

The implications of this work are fundamental to our understanding of the universe and its history. If the universe is not expanding and the FRW model is invalid, there was no big bang and the general cosmological model will have to be replaced with another approach. There are alternative cosmological models that can explain major features of the universe such as large scale structure, the light element abundances and the cosmic background radiation without a big bang or an expanding universe. For example, plasma cosmology, which assumes an evolving universe without an origin in time has provided such explanations and accurate predictions of new phenomena.[16,17,18] Such alternatives have received additional justification in recent years as contradictions with the big bang model have accumulated in many areas. These include the failure of predictions of the abundance of lithium[19], the non-Gaussianity of the CMB anisotropies[20] and the failure to predict very large voids in the distribution of galaxies[21].

In addition, if $d=cz/H$, there is no absolute limit to the size of the observable universe, nor any limit to how far back in time we can observe. The observable universe would, in principle, be unlimited in both space and time. Specifically, objects in the universe could be far older than the 14Gy allowed by the big bang hypothesis. For example, if $d=cz/H$ the galaxies in the HUDF observed at z=6 are 24 Gpc away and are observed as they were 72 Gy ago. This in turn means that at least some galaxies existing today are that old.

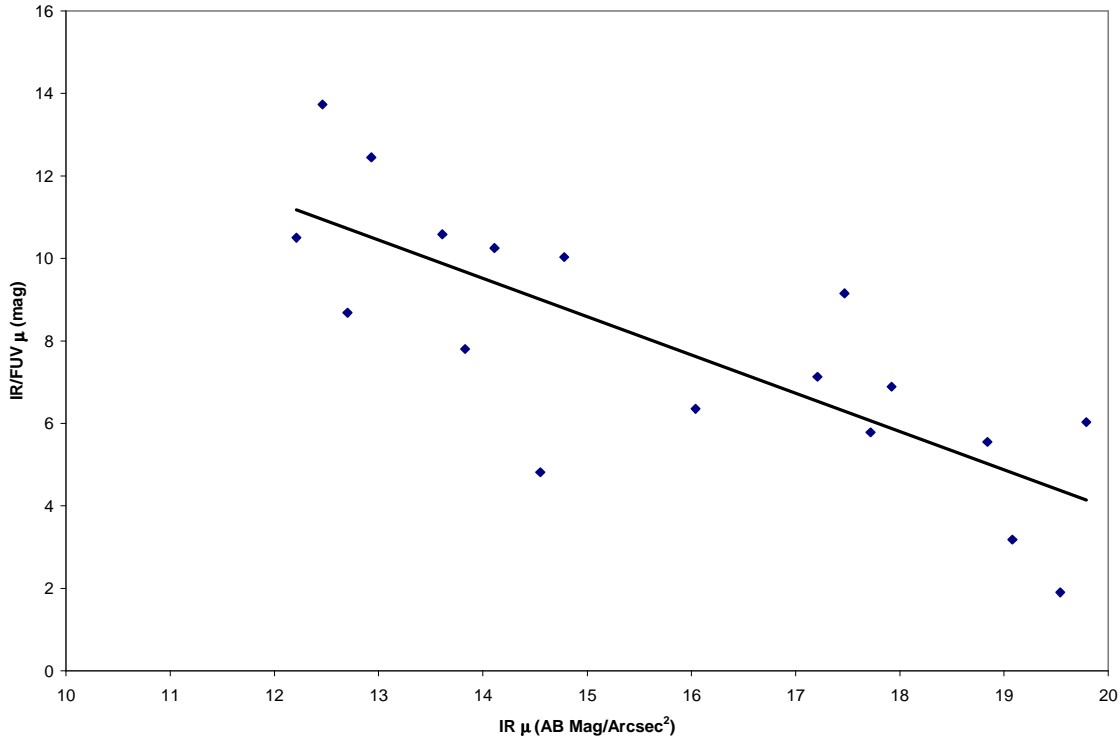

**FIGURE 9.** IR/FUV surface brightness ratio for nine ULIRG galaxies. All those with IR $\mu$ brighter than 19 show strong absorption.

Finally, the surface brightness data implies that the Hubble relation must be caused by some non-geometric process. One possibility is loss of energy from photons due to some previously unknown process. A linear relationship between z and d is mathematically equivalent to each photon loosing a fixed quantum of energy for each wavelength traveled. Spherically the quantum of energy lost would be that of a photon with a wavelength equal to the Hubble distance, 4,000 Mpc.

Give the implications of this work, it's crucial to see if these conclusion are consistent with other observations and how sensitive they are to the model hypothesized. A full investigation of this question requires additional work, as detailed in section 5, but a start can be made here. First, how sensitive are the conclusions to the assumption that d=cz/H? We can look briefly at the comparison of the z=6 sample using instead the assumption that d=cln(1+z)/H, which is a commonly-used formula for non-expanding cosmologies.

With this assumption, the distance at z=6 is 0.32 of the distance assuming d=cz/H, so absolute magnitudes for the same galaxies are increased by 2.44 and the low-size limit is reduced to 2.3 kpc—although in fact none of the galaxies in the sample is, with this assumed relationship, less than 3.4 kpc. We examine samples with 25.5<M<26.5. A comparison FUV sample at z=0.06-0.07 has a low-size cut off of 2.3 kpc and contains 124 galaxies. The average surface brightness in this sample is 24.60 as compared with 24.22 in the high-z sample of 25 galaxies, a difference of only 0.38 magnitudes. While the low-z sample is dimmer in this analysis than for the d=cz/H assumption, the low-size cut-off is also smaller, resulting in little change in average surface brightness. This individual comparison again implies a good agreement with the prediction, common to all non-expanding models, of no change of surface brightness with redshift. A more through analysis is clearly desirable, but at first glance we see that the conclusion that the universe in not expanding is relatively insensitive to the choice of d-z relationship. By the same reasoning, surface brightness data alone is probably not a good way to distinguish between d-z relationships in non-expanding models.

A second question is whether these conclusions are compatible with observations of Type Ia supernovae, which have also been used to measure the geometry of the universe. Many authors have claimed, in particular, that the decay curves of supernovae prove the expansion of the universe, as they are longer at higher redshift, implying a stretching of the duration of the event, not just a simple redshifting of the light[22]. However, as Brynjolfsson notes[23], there is circularity in this argument because observations have demonstrated that, at a given redshift, the time constant of a SNIa decay cure is linearly proportional to the peak absolute magnitude of the supernova[24].

Consider a given supernova with apparent magnitude m, redshift z and observed duration t. In the FRW interpretation the intrinsic duration is t/z+1 and the absolute magnitude

$$M = m - 5\log d - 2.5\log(z+1) \qquad (7)$$

where the z+1 factor is due to the decrease in flux caused by the stretching of the duration of the pulse. We also have that

$$M = M_o - 2.5\log t + 2.5\log(z+1) \qquad (8)$$

So, combining eqs.(7) and (8), we have

$$m = M_o - 2.5\log t + 5\log d + 5\log(z+1) \qquad (9)$$

In the non-expanding model, the intrinsic duration is t and the absolute magnitude is

$$M = m - 5\log d' \qquad (10)$$

where d' is the distance in the non-expanding model, which, at low z, equals (z+1)d. Thus,

$$M = m - 5\log d - 5\log(z+1) \qquad (11)$$

In this case,

$$M = M_o - 2.5\log t \qquad (12)$$

and therefore the expression for m formed by combing (11) and (12) is identical with (9) and the FRW and non-expanding models are indistinguishable. A more detailed comparison is planned using available SNIa data.

In addition, if the relationship d=cz/H is assumed, the variation of apparent magnitude of a given source with redshift is extremely similar for the non-expanding universe and for the "consensus" FRW model. Over a range in z from 1-6 the difference in the two models predictions is less than 0.21 magnitudes at all z.

## 5. CONCLUSIONS AND FUTURE WORK

The data from HUDF and GALEX give a clear answer to the surface brightness test of the expansion of the universe. The data is clearly compatible with the non-expanding hypothesis and clearly incompatible with the expanding hypothesis, even with evolution. The universe, therefore, is not expanding. Surface brightness is independent of redshift to well within narrow statistical uncertainties, while the FRW evolutionary hypothesis requires that high-z galaxies have FUV surface brightness that are more than an order of magnitude outside the entire range of low-z galaxies and which appear to be physically impossible.

There are a number of directions for further work to confirm and expand these conclusions. The analysis presented here comparing non-expanding models with the surface brightness data needs to be expanded to include other redshift ranges, filling in the entire range from z=0-6. The surface brightness relation has to be examined in other at-galaxy wavelength ranges in the optical. A more thorough sensitivity analysis must be undertaken to see how well alternative d-z relationships fit the data at the full range of redshifts and wavelengths examined.

Since the FRW evolutionary hypothesis can only be ruled out by excluding the existence of the extremely high-surface-brightness FUV galaxies at high-z, a number of efforts are needed. To find the brightest galaxies at low-z the entire GALEX data base, not just those with SDSS redshift, can searched for comparison galaxies. Similarly, a more thorough search of all HUDF, HDF and other high-z data bases is required to find the brightest galaxies. In addition, theoretical models need to be examined to see if any can allow, or rule out, ultra-bright UV galaxies.

The implications of this work require the re-examination of other data as well. For one thing, the SNIa data for individual galaxies can be re-analyzed on the basis of a non-expanding universe model, although for the theoretical reason mentioned in section 4, it is not anticipated that the SNIa data alone can distinguish between expanding and non-expanding models. In addition, the non-expanding model implies the existence of much older galaxies at all epochs than does the FRW model. While age-determination of galaxies is fraught with analytical ambiguities, studies need to be undertaken to test if galaxies much older than 14 Gy exist today, or similarly old galaxies at high-z.

It is important to note as well that the data bases that this work relies on can be used for other tests that bear on the geometry of the universe. For example, the data can be used to indicate the change in galaxy density with z, which in turn depends of the z-d relationship, and thus can potentially distinguish such models.

Finally, if the universe is not expanding and the Hubble relationship is due to some other phenomenon, it seems likely that such a redshift-distance relationship could be detected on earth with sufficiently sensitive instruments. Experiments, possibly with a modification of the Laser Interferometer Gravity Observatory, could in principle detect a change in wavelength of light and thus determine experimentally the origin of the Hubble relationship.

## ACKNOWLEDGMENTS

I want to thank Dr. Kirk Borne, (George Mason University), Dr. Timothy Eastman (Plasmas International) and Simone Cook, (student intern at NASA Goddard SFC and Spellman College) for their help in preparing some of the data bases used in this study.